\documentclass[preprint,12pt]{elsarticle}
\usepackage{graphicx}
\usepackage{dcolumn}
\usepackage{bm}
\usepackage{amsfonts}
\usepackage{color}
\usepackage{epsfig}
\usepackage{times}
\usepackage{booktabs} 
\usepackage{epstopdf}
\usepackage{multirow}
\usepackage{float}
\usepackage{amsmath}
\usepackage{algpseudocode}
\usepackage{threeparttable}
\usepackage{algpseudocode}

\usepackage{ragged2e}
\usepackage{ulem}

\usepackage{orcidlink}
\usepackage{cases}
\usepackage[utf8]{inputenc}
\usepackage[T2A]{fontenc}
\usepackage[greek,russian,spanish,english]{babel}
\usepackage[T1]{fontenc}

\usepackage{appendix}

\usepackage{threeparttablex}

\usepackage{colortbl}  
\usepackage{array}

\usepackage[section]{placeins} 

\usepackage{amssymb}

\usepackage{cleveref}

\journal{xxx}

\begin{document}

\begin{frontmatter}

\title{A Quick Framework for Evaluating Worst Robustness of Complex Networks}

\author[fn1]{Wenjun Jiang\orcidlink{0000-0002-2244-5196}}
\author[fn1]{Peiyan Li\orcidlink{0009-0004-4872-0477}}
\author[fn2]{Tianlong Fan\corref{cor1}\orcidlink{0000-0002-9456-6819}}
\ead{tianlong.fan@ustc.edu.cn}
\author[fn1]{Ting Li\orcidlink{0009-0009-2105-5602}}
\author[fn1]{Chuan-fu Zhang\corref{cor1}}
\ead{zhangchf9@mail.sysu.edu.cn}
\author[fn1]{Tao Zhang}
\author[fn1]{Zong-fu Luo\corref{cor1}\orcidlink{0000-0003-4871-4216}}
\ead{luozf@mail.sysu.edu.cn, 
 +86-020-84114504}
\affiliation[fn1]{organization={School of Systems Science and Engineering, Sun Yat-sen University},
city={Guangzhou},
postcode={510275},
country={China}}
\affiliation[fn2]{organization={School of Cyber Science and Technology, University of Science and Technology of China},
city={Hefei},
postcode={230026},
country={China}}

\cortext[cor1]{Corresponding author}


\begin{abstract}
Robustness is pivotal for comprehending, designing, optimizing, and rehabilitating networks, with simulation attacks being the prevailing evaluation method. Simulation attacks are often time-consuming or even impractical, however, a more crucial yet persistently overlooked drawback is that any attack strategy merely provides a potential paradigm of disintegration. The key concern is: in the worst-case scenario or facing the most severe attacks, what is the limit of robustness, referred to as ``Worst Robustness'', for a given system? Understanding a system's worst robustness is imperative for grasping its reliability limits, accurately evaluating protective capabilities, and determining associated design and security maintenance costs. To address these challenges, we introduce the concept of Most Destruction Attack (MDA) based on the idea of knowledge stacking. MDA is employed to assess the worst robustness of networks, followed by the application of an adapted CNN algorithm for rapid worst robustness prediction. We establish the logical validity of MDA and highlight the exceptional performance of the adapted CNN algorithm in predicting the worst robustness across diverse network topologies, encompassing both model and empirical networks. This Worst Robustness Evaluation (WRE) framework has remarkable scalability, accommodating various attack strategies, be they existing or prospective, and upgrading predictive capabilities with more powerful machine learning algorithms, the WRE framework can further enhance its performance.
\end{abstract}


\end{frontmatter}


\section{Introduction}

Robustness \cite{albert2000error,lordan2019exact,wang2022robustness}, specifically connectivity robustness, refers to the ability of a networked system to maintain structural integrity when facing random component failures or malicious external attacks. In this context, maintaining a certain level of structural integrity is considered a necessary condition for the network to sustain its normal operation. For nearly any such system, accurately assessing their robustness is a fundamental concern in the design of their topological architecture \cite{sydney2010characterising, peng2016optimal, safaei2020robustness}, vulnerability analysis \cite{liu2017recognition, mishkovski2011vulnerability}, and functional maintenance \cite{cohen2010complex}. Grounded in percolation theory \cite{cohen2000resilience}, a value sequence representing a specific characteristic of the residual network is commonly employed to assess the robustness of a network during failure or attack. These assessment methods can be categorized into four types \cite{jiang2023comprehensive}, with the most prevalent being the sequence of changes in the size of the giant connected component (GCC) of the network. This sequence, simple and direct, reflects the scale of the core network that continues to function amidst failure or attack.

In practice, a more crucial and pressing real-world question is how a network performs in its worst-case scenario in terms of robustness. Specifically, this refers to the limit of robustness exhibited by a system when subjected to the most destructive attacks. We term the lower bound of this robustness as ``Worst Robustness’’. Only by comprehending the worst robustness of a system can we more accurately assess its defensive capabilities against unknown attacks and determine the costs associated with the design and maintenance of its security \cite{paul2004optimization}. However, existing research on robustness often relies on one or more potential attack paradigms. Even if these attacks exhibit considerable destructiveness, they may not represent the system's worst robustness. This holds true regardless of whether the attack strategies are based on node centrality \cite{albert2000error,lu2016vital,fan2021characterizing,wang2024identifying,wang2021method,wandelt2023measuring}, percolation \cite{morone2015influence,morone2016collective,qiu2021identifying}, cascading failures \cite{motter2002cascade,lai2004attacks}, iterative adaptive methods \cite{holme2002attack, ren2019generalized}, heuristic strategies \cite{zdeborova2016fast, deng2016optimal, mugisha2016identifying}, or machine learning methods \cite{li2019new, fan2020finding, grassia2021machine}. On the other hand, although percolation models on random graphs can provide precise expressions for the position of GCC appearance and other topological statistics, irrespective of whether the network's degree distribution is non-Poissonian \cite{newman2001random} or completely general \cite{callaway2000network}, they often focus only on the vicinity of the phase transition. Such models cannot evaluate the overall performance of network robustness throughout the entire process and cannot address the question of worst robustness.

While studies on network dismantling \cite{braunstein2016network,wandelt2021estimation} and optimal attacks \cite{deng2016optimal, peng2023unveiling} are also fundamentally pursuing this limit state, existing knowledge fails to provide a singular or analytical solution to address this issue, aside from enumerating all possibilities. The problem of worst robustness can be reframed as the question of how to most effectively dismantle a network to induce the fastest decrease in its GCC size. Directly seeking its exact solution as an NP-hard problem is impractical, particularly for large-scale networks. Therefore, finding better ways to capture its suboptimal solutions becomes an urgent and practically significant issue. Although existing dismantling strategies may not yield optimal solutions, they might demonstrate the ability to inflict the most severe damage at a certain stage or in specific sections. Motivated by this insight and based on the concept of knowledge stacking, we can extract the most destructive parts of each known strategy and assemble them into a complete, continuous strategy with the maximum known destructive power throughout the entire process. We define this strategy as the Most Destruction Attack (MDA),
as shown in Figure \ref{fig:schematic}. According to this definition, the damage caused by any single or combined attack strategy cannot exceed that of MDA. Therefore, under the conditions of MDA, the network's exhibited robustness can be defined as the worst robustness. Given any networked system, under any type of component failure (such as nodes or edges), the worst robustness can be captured through MDA simulation. Thus, this approach offers a universally applicable and available suboptimal solution.

To further address the time-consuming challenges associated with evaluating the worst robustness of networks using MDA, we introduce an adapted and well-validated deep learning algorithm, known as adapted CNN-SPP\cite{jiang2023comprehensive}. Through comprehensive training of the adapted CNN-SPP with the results derived from MDA, we achieve a rapid and accurate assessment of the network's worst robustness.

In the context of connectivity robustness under node failure scenarios, this paper addresses the problem of the network's worst robustness and introduces the concept of the MDA to facilitate a practical and feasible assessment of the network's worst robustness. We establish the rationality validation of the MDA and subsequently enhance the efficiency of this assessment process by invoking an adapted CNN-SPP model. The results on networks with diverse topological features demonstrate the effectiveness and efficiency of this framework. Until the theoretical optimum solution is identified, the Worst Robustness Evaluation Framework (WRE framework) will be imperative in guiding system security design and protection cost budgets. This holds true for both infrastructure networks and adversarial networks, carrying significant practical implications. 

Finally, the paper discusses the extensibility of the WRE framework in terms of approximating the optimal solution and enhancing predictive capabilities. The performance boundary issues addressed by the WRE framework have broad generality. In addition to scenarios involving edge failures, the WRE framework applies to the rapid assessment of the network's optimal control robustness. In this case, one only needs to replace attack strategies and GCC size change sequences with control strategies \cite{fan2021characterizing,yu2013synchronization, amani2017finding} and the corresponding network synchronization capability evaluation metrics \cite{pirani2015smallest, liu2018optimizing}. Moreover, this paradigm holds significant potential in the realm of crucial node identification \cite{fan2021characterizing}.

The remaining content is organized as follows. The remainder of this article is organized as follows. Sec. \ref{ourframework} presents the proposed worst robustness evaluation framework, Sec. \ref{results} showcases our results and analysis. Finally, Sec. \ref{discussion} discusses the study and its extensions in terms of performance and applications.

\section{Worst Robustness Evaluation Framework}\label{ourframework}

We propose a worst robustness evaluation framework (WRE framework) to achieve a rapid and feasible evaluation of the Worst Robustness of networks. This framework primarily consists of two components. The first part aims to capture the network's Worst Robustness. Specifically, we introduce the concept of the Most Destruction Attack (MDA). When the network undergoes an MDA, the exhibited robustness defines the Worst Robustness. The second part of the framework addresses the efficiency issue in the assessment process. Since the capturing process in the first part requires numerous time-consuming simulations, it tends to be slow. To overcome this, we employ a convolutional neural network with spatial pyramid pooling (CNN-SPP). After appropriate training, it enables a rapid and accurate evaluation of the Worst Robustness in networks.

\subsection{Most Destruction Attack}

A network can be subjected to various attack modes, including the removal of a single node along with its adjacent edges, a specific edge, a set of multiple edges, or a collection of multiple nodes with their adjacent edges. The connectivity destruction, denoted as $D(p)$, occurs after removing nodes with a proportion $p$ (including all their adjacent edges), representing the loss of connectivity compared to the initial state, and is defined as $D(p)=G(0)-G(p)$. Here, $G(p)$ and $G(0)$ respectively denote the relative size of the GCC in the remaining network after removing nodes with proportion $p$ and the original network. If there exists an attack strategy that maximizes $D(p)$ at any removing proportion $p$, it is referred to as the Most Destruction Attack (MDA). In this scenario, the network's connectivity destruction, denoted as $D_{MDA}$, is defined as

\begin{equation}
    D_{MDA} = \frac{1}{N} \sum_{p=1/N}^1 {\rm max} \; D(p) = \frac{1}{N} \sum_{p=1/N}^1 {\rm max} [G(0)-G(p)],
    \label{equ:D_MDA}
\end{equation}

\noindent where $N$ is the number of nodes. Attacking the network in the MDA manner until no nodes remain, the resulting attack curve is termed the Most Destructive Attack Curve (MDA curve). In this context, the term ``attack curve'' refers to the sequence of values representing the relative size of the GCC in the remaining network at each $p$ during the attack process.

\subsection{Worst Robustness}

When the network undergoes MDA, the robustness exhibited by the network is referred to as Worst Robustness, denoted as $R_W$,

\begin{equation}
    R_{W} =1- D_{MDA}=1-\frac{1}{N} \sum_{p=1/N}^1 D_{MDA}(p),
    \label{equ:RW_1}
\end{equation}
 
\noindent where $D_{MDA}(p)$ denotes the relative size of the GCC of the network at $p$ when subjected to the MDA. The subsequent sections illustrate the discussion using the one-by-one node removal attack mode as an example. It's worth noting that other attack modes would not lead to different conclusions.

\subsection{MDA Based on Knowledge Stacking}

Given that theoretically capturing MDA within feasible time constraints is currently impractical, here we present a practically actionable method. This method stacks existing dismantling knowledge, concatenating measures that exhibit the maximum destructive effect at each step to form a complete dismantling solution. We validate the rationality of this approach.

Specifically, we consider $q$ attack strategies. At each removal step, they offer $q$ choices. We select the node that inflicts the maximum damage on the network at each step. It is worth noting that in practice, $ {\rm max} [G(0)-G(p)]$ is equivalent to $ {\rm min} [G(p)]$. This process continues until all nodes are removed one by one. Subsequently, we concatenate the sequentially selected nodes to obtain a practical MDA scheme. Therefore, at the practical level, Equation \ref{equ:RW_1} can be transformed into:

\begin{equation}
    R_{W} = \frac{1}{N} \sum_{p=1/N}^1 {\rm min} [G_{S_1}(p),G_{S_2}(p),...,G_{S_q}(p)],
    \label{equ:RW_real}
\end{equation}

\noindent here, $G_{S_q}(p)$ represents the relative size of the GCC of the remaining network when nodes are removed according to strategy $S_q$ at proportion $p$. This MDA scheme serves as a practical approximation of the theoretically optimal scheme, with the disparity between them depending on the number and quality of selected attack strategies: the stronger the attack capabilities of each strategy and the greater their number, the closer the resulting MDA scheme will be to the theoretically optimal one. It is noteworthy that sometimes different attack strategies may offer the same choices. In such cases, we should prioritize selecting nodes that have not been previously chosen to avoid multiple occurrences of the same node in the final MDA scheme.

\begin{figure} [htbp]
\centering
\includegraphics[width=12.8cm]{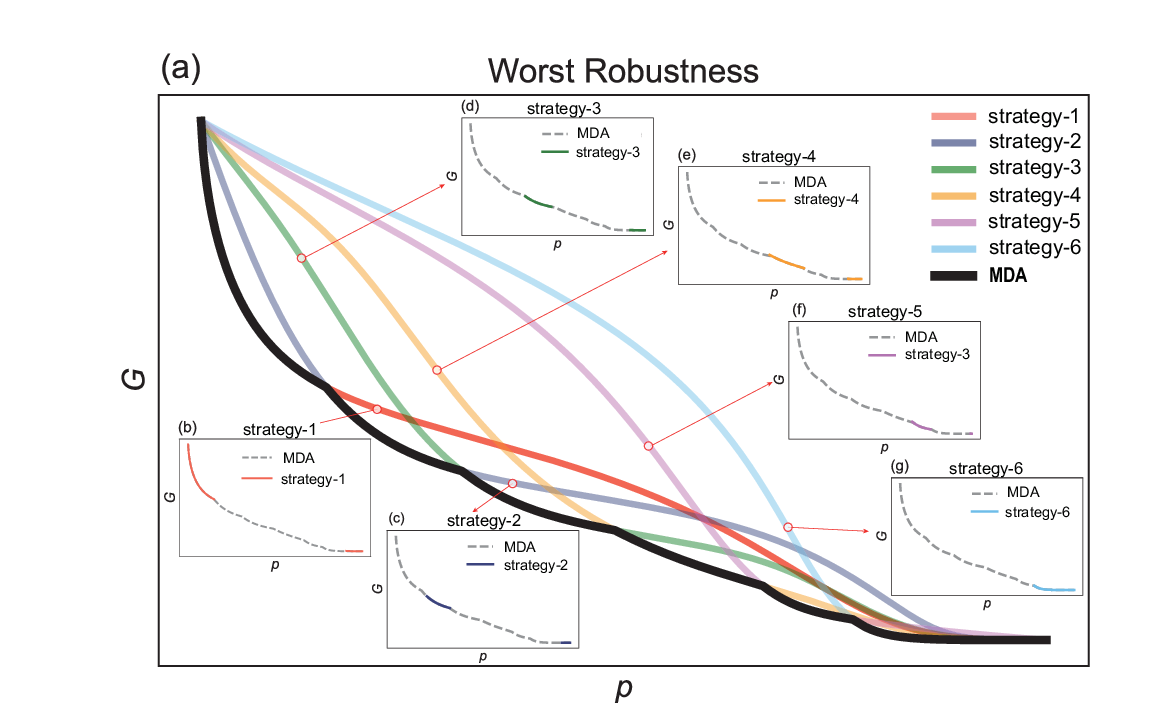}

    \caption{The schematic of the Most Destruction Attack (MDA) and the worst robustness of the network. Here, $p$ and $G$ represent the node removal ratio and the relative size of the largest connected component, respectively. The MDA is obtained by extracting the most destructive portions from each attack paradigm and combining them. The MDA demonstrates the most severe damage that a network may undergo, guided by existing knowledge, and the worst robustness exhibited under such circumstances.}
    	\label{fig:schematic}
\end{figure}

Figure \ref{fig:schematic} illustrates the capturing process of such an MDA scheme. The main panel (a) displays the capturing process of MDA, which exhibits maximum destructive capability at any position, or in other words, minimizes the relative size of the GCC. It is constructed by stacking multiple alternative strategies and selecting the optimal solution for each position. Insets (b) to (g) represent six alternative strategies, where the dashed lines denote the MDA curve, and the solid segments represent the contribution of each strategy to the MDA. Attacking the network based on this MDA scheme captures the robustness of the network, which represents the worst-case robustness of the network.

In this study, we take $q=8$ centrality metrics as representatives of existing network attack strategies, namely: Degree \cite{iyer2013attack}, H-index \cite{hirsch2005index}, Coreness \cite{kitsak2010identification}, Closeness \cite{sabidussi1966centrality}, Betweenness \cite{freeman1977set}, Eigenvector \cite{bonacich1987power}, Pagerank \cite{brin1998anatomy}, and Cycle ratio \cite{fan2021characterizing}. Detailed descriptions of these strategies can be found in Sec. 1 of \nameref{sec:methods}. Note that the static versions of these metrics are used here, where centrality values for each node are computed only once for each metric. Figure \ref{fig:sample} in \nameref{sec:appendix} illustrates an example of MDA curve for an empirical miscellaneous network \cite{nr}, demonstrating that the MDA approach exhibits maximum destructive capability at any position. The top-left panel contrasts the attack curves derived from the eight centrality metrics with the MDA. The remaining panels display the contributions of each metric to the MDA.

\subsection{Rationality Verification of MDA}

An issue that must be addressed is the extent of rationality achieved by such stacking methods in generating MDA solutions. In other words, how low is the rate of node non-repetition in this MDA solution? In the most ideal case, each node should be removed and can only be removed once, meaning that all nodes on the MDA curve appear only once. To investigate this, we experimented to validate the rationality of MDA. Specifically, we first defined a metric, termed Maximum Rationality (MR), which refers to the maximum proportion of nodes appearing only once in the MDA curve. The detailed calculation process of MR is provided in Sec. 2 of \nameref{sec:methods}.

\begin{figure} [h]
\centering
\includegraphics[width=\linewidth]{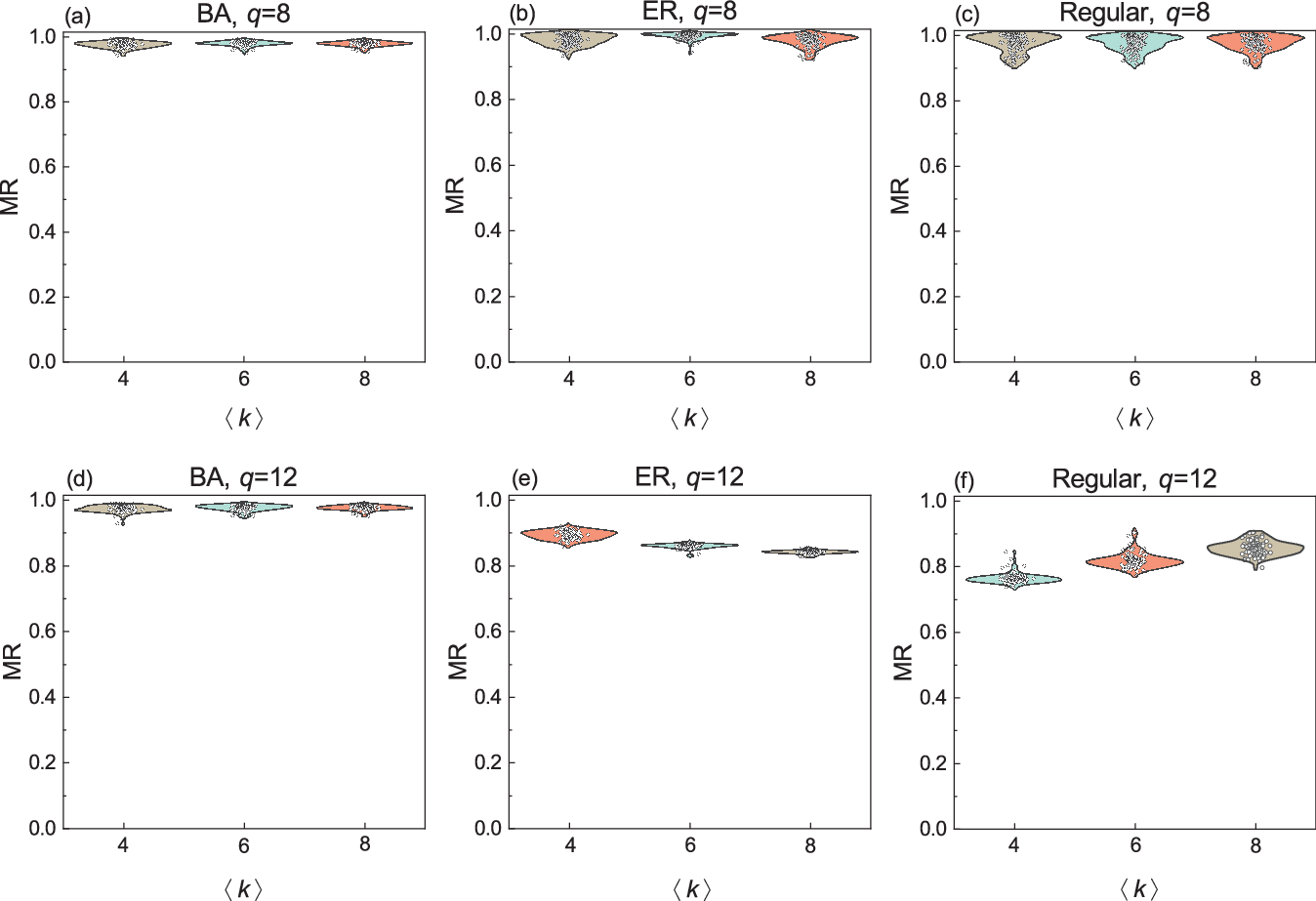}
\caption{Maximum rationality of three types of synthetic networks. Here, $q$ represents the number of centrality metrics considered to capture MDA, the $x$-axis denotes the average degree $\langle k\rangle$ of the network, and the $y$-axis represents the value of MR. Each point in the panel represents a network, with each parameter comprising 100 instances of networks with a node count of 5000.}
	\label{fig:Rat_Ver}
\end{figure}

Figure \ref{fig:Rat_Ver} presents the statistical results of maximum rationality in three types of synthetic networks, including Barabási–Albert model (BA) networks \cite{barabasi1999emergence}, Erdős–Rényi model (ER) networks \cite{erdHos1960evolution}, and Regular networks \cite{erdHos1960evolution}, representing heterogeneous, homogeneous, and regular networks, respectively. When $q=8$, regardless of the network's density, the MR of all three networks approaches or equals 1 with no significant differences between them, and BA networks exhibit the smallest variance. As $q$ increases to 12, the MR of BA networks remains largely unaffected, while that of ER networks and Regular networks decreases overall. Additionally, as the network becomes denser, the MR of ER networks slightly decreases, while that of Regular networks slightly increases. This phenomenon primarily arises from the significant structural diversity among nodes in heterogeneous networks, where the values of centrality metrics often exhibit considerable variations. As a result, different metrics can effectively differentiate nodes, and their rankings may be similar, thus stabilizing the MR of BA networks near 1, regardless of the average degree $\langle k \rangle$ and the number of centrality metrics $q$ considered during MDA computation.

Conversely, for homogeneous networks, especially Regular networks, most nodes have similar centrality metric values due to their comparable topological features, making it difficult for metrics to effectively distinguish the importance of nodes. This results in significantly varied rankings, leading to multiple occurrences of the same node at different positions in the MDA curve, with the degree of repetition intensifying with the increase in alternative strategies, consequently yielding a lower MR. Furthermore, as $\langle k \rangle$ increases, the degree values of nodes in ER networks exhibit a broader range and a more dispersed distribution, as shown in Figure \ref{fig:ERdistr} in \nameref{sec:appendix}, reducing the substitutability of nodes at specific positions in the MDA curve and causing a downward trend in MR. Conversely, in Regular networks, where nodes share identical topological characteristics, the substitutability of nodes at specific positions in the MDA curve increases, resulting in an upward trend in MR.

Tables \ref{tab:A1} and \ref{tab:A2} in \nameref{sec:appendix} present the maximum, minimum, mean MR values, as well as the right-tail distribution for different synthetic networks and their parameter configurations. In summary, the overall MR of the three networks is relatively high, particularly in the case of BA networks, which exhibit a very high MR that remains largely unaffected by network density and the number of alternative strategies considered during MDA computation. These results demonstrate the rationality of our MDA approach and suggest the potential value of the WRE Framework in WR assessment. This also suggests that the number of alternative strategies in MDA should not be more, but rather a combination of a few high-quality strategies.

\subsection{Quick Evaluator Based on CNN}

The process of capturing MDA requires stacking the results of multiple attack schemes, all of which necessitate time-consuming simulations, posing a significant obstacle in practical applications. To address this challenge, we employed an adapted and performance-validated deep learning algorithm, adapted CNN-SPP. By extensively training this CNN-SPP model using MDA results, we can rapidly and accurately assess the MDA of networks, thus deriving their Worst Robustness. 

\begin{figure} [htbp]
\centering
\includegraphics[width=1.9\linewidth]{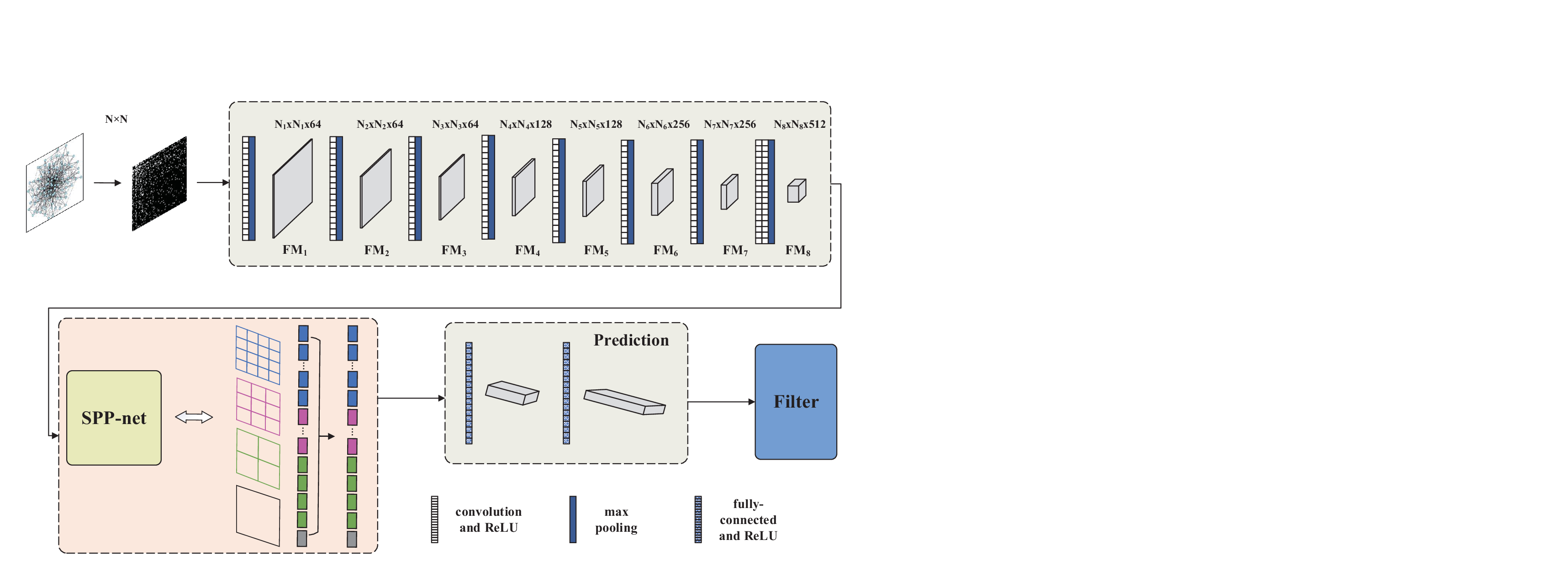}
\caption{Architecture of the proposed quick evaluator based on CNN. It comprises five main components, namely the network's input section, convolutional set, SPP-net, two fully connected layers, and a filtering section. The convolutional set consists of eight convolutional blocks, each typically comprising one or two convolutional layers, a rectified linear unit (ReLU) activation function, and a max-pooling layer. The size of the $i$-th feature map (FM) FM$_i$ is denoted as $N_i= \lceil \frac{N}{2^{i}} \rceil$, for $i=1,2,...,8$, where $N$ represents the number of nodes in the input network.}
	\label{fig:cnn}
\end{figure}

CNNs are tailored for the processing of grid-structured data, they incorporate the feature extraction function directly into a multilayer perceptron via structural rearrangement and weight reduction \cite{simonyan2014very}. On the other hand, the adjacency matrix of the network can be directly viewed as a binary image: where 1 indicates the existence of the corresponding edge, and 0 indicates absence. Therefore, CNNs excel in processing networks, enabling instantaneous evaluation of robustness post-training, and demonstrating outstanding generalization capabilities.

The architecture of this CNN model is illustrated in Figure ~\ref{fig:cnn}. Specifically, it consists of five main components: the network's input section, eight convolutional blocks, the spatial pyramid pooling layer, two fully connected layers, and a filtering mechanism designed to handle irrational values post-prediction.

The input layer efficiently transmits input data, represented by the adjacency matrix of a network, for subsequent feature extraction and learning. The convolutional layer, serving as the foundation of CNN, achieves local feature detection through convolutional operations, with each kernel capturing distinct local features to form abstract high-level representations. To introduce non-linearity, the Rectified Linear Unit (ReLU) activation function \cite{nair2010rectified} is applied after the convolutional layer, enhancing the network's adaptability to complex image patterns and increasing its expressive power. Stacking multiple groups of convolutional layers and ReLU activation functions allows the network to progressively learn higher-level and more abstract features, thereby enhancing its capacity for processing input images. The max pooling layer is utilized for downsampling feature maps (FM), reducing spatial dimensions, minimizing computational costs, and improving the network's robustness to translation invariance, with max pooling chosen to retain the most significant features. The parameters of each group of convolutional layers are detailed in Table \ref{tab1}.


\begin{table}[t]
\caption{Parameters of the eight groups of convolutional layers}
\centering
\begin{tabular}{ccccc}
\toprule
\textbf{Group}                   & \textbf{Layer}     & \textbf{Kernel} & \textbf{Stride} & \begin{tabular}[c]{@{}c@{}}\textbf{Output}\\ \textbf{channel}\end{tabular} \\ 
\midrule
\multirow{2}{*}{Group1} & Conv3-64  & 3×3    & 1      & 64                                                       \\
                        & Maxpool   & 2×2    & 2      & 64                                                       \\
\multirow{2}{*}{Group2} & Conv3-64  & 3×3    & 1      & 64                                                       \\
                        & Maxpool   & 2×2    & 2      & 64                                                       \\
\multirow{2}{*}{Group3} & Conv5-64  & 5×5    & 1      & 64                                                       \\
                        & Maxpool   & 2×2    & 2      & 64                                                       \\
\multirow{2}{*}{Group4} & Conv3-128 & 3×3    & 1      & 128                                                      \\
                        & Maxpool   & 2×2    & 2      & 128                                                      \\
\multirow{2}{*}{Group5} & Conv3-128 & 3×3    & 1      & 128                                                      \\
                        & Maxpool   & 2×2    & 2      & 128                                                      \\
\multirow{2}{*}{Group6} & Conv3-256 & 3×3    & 1      & 256                                                      \\
                        & Maxpool   & 2×2    & 2      & 256                                                      \\
\multirow{2}{*}{Group7} & Conv3-256 & 3×3    & 1      & 256                                                      \\
                        & Maxpool   & 2×2    & 2      & 256                                                      \\
\multirow{3}{*}{Group8} & Conv3-512 & 3×3    & 1      & 512                                                      \\
                        & Conv3-512 & 3×3    & 1      & 512                                                      \\
                        & Maxpool   & 2×2    & 2      & 512                                                      \\ \hline
\end{tabular}
\label{tab1}
\end{table}

The fully connected layer serves to map features acquired from the convolutional layers to the final output categories, necessitating a consistent input size. Addressing this requirement, the SPP-net (Spatial Pyramid Pooling Network) \cite{he2015spatial} acts as an intermediary, generating a fixed-length representation regardless of input image size by capturing salient responses across various scales and producing a resultant vector. By combining SPP-net with the fully-connected layer, our model can handle inputs of varying dimensions. The SPP-net allows handling objects of different scales and sizes, thereby enhancing object recognition robustness and accuracy. It partitions the image into varying levels, consolidating local features within these partitions.

In our model, we integrate a 4-level pyramid from SPP-net, incorporating sizes of {4 × 4, 3 × 3, 2 × 2, 1 × 1}, aiming to generate a fixed-length vector through pooling feature maps. This pyramid configuration, depicted in Figure \ref{fig:cnn}, comprises $16 + 9 + 4 + 1 = 30$ bins. Following the last convolutional block, the model boasts 512 channels (as listed in Table \ref{tab1}), yielding a total of $30 \times 512 = 15,360$ fixed-length vectors. This transformation facilitates effective data processing by the fully connected layer to capture intricate relationships and learn higher-level features.

After obtaining the prediction results, it is common to encounter values that deviate unreasonably from the expected range. The purpose of the filter is to rectify these anomalies, ensuring that the final prediction results are logical and interpretable. This may involve removing outliers, smoothing prediction outcomes. During simulated attacks, the relative size of any GCC in the MDA curve falls within the interval $[0,1]$, and the entire curve is monotonically non-increasing. To address this issue, we introduce a simple modified filter \cite{lou2021convolutional} consisting of two parts: the first part eliminates superdomain data, as shown in Equation \ref{eq:Gi1}; the second part ensures the curve exhibits monotonically non-increasing features, as represented by Equation \ref{eq:Gi2}.




\begin{equation}\label{eq:Gi1}
    G_{MDA}(i) = \left\{ \begin{aligned}
  & 1,G_{MDA}(i) > 1 \\ 
 & 0, G_{MDA}(i)< 0,\\ 
 & G_{MDA}(i), others \\ 
\end{aligned} \right.
\end{equation}

\begin{equation}\label{eq:Gi2}
G_{MDA}(j) = G_{MDA}(i-1) - \frac{j-(i-1)}{k-(i-1)} \cdot [G_{MDA}(i-1) - G_{MDA}(k) ], i \leq j < k\leq N.
\end{equation}

In Equation \ref{eq:Gi2}, the MDA sequence $[G_{MDA}(1), G_{MDA}(2), ..., G_{MDA}(N)]$ is systematically evaluated to ascertain whether each value violates the monotonic non-decreasing condition. Upon detecting a subsequence $[G_{MDA}(i), ..., G_{MDA}(j)]$, where $i \leq j$, that fails to meet this condition, and where the subsequent $k$th value satisfies $G_{MDA}(k) \leq G_{MDA}(i-1)$, all values from $G_{MDA}(i)$ to just before $G_{MDA}(k)$ are reassigned according to Equation \ref{eq:Gi2}.

The main tasks of loss function include measuring the model's performance on training data, providing an optimizable objective for the model and reflecting specific task goals. By quantifying the differences between model predictions and actual targets, the loss function guides the adjustment of model parameters, allowing the model to better fit the training data and providing effective feedback throughout the training process. The Mean Squared Error (MSE) function is a commonly used loss function in machine learning and deep learning, Its form is as follows: 

\begin{equation}\label{eq:L2loss}
J_{MSE}=\frac{1}{H}\sum_{i=1}^{H} \parallel y_i-\hat{y}_i \parallel^2,
\end{equation}
and $\parallel \cdot \parallel$ is the Euclidean norm, $y_i$ and $\hat{y}_i$ are the $i$th element of the predicted vector and the simulated vector respectively, and $H$ is the length of them. The objective of training the model is to minimize the $J_{MSE}$ as much as possible, with its minimum value of 0 indicating that the predicted results are identical to the simulation.

\subsection{Training of the Quick Evaluator}

To train the quick evaluator, we constructed a corresponding training dataset. This dataset comprises the adjacency matrix of a network as the input object, with the label being the curve generated by the MDA encompassing eight attack strategies. While the SPP-net offers two input modes—single-size training and multi-size training—we have opted for the former for dataset convenience.

The model training comprises two stages, with the first stage utilizing solely synthetic network data, and the second stage augmenting it with additional empirical network data. We set the size of the training networks to $N=1000$. Each specific network model and mean degree encompass 1000 network instances, with 800 used for training, 100 for cross-validation, and the remaining 100 for testing. Thus, there are a total of 4 model networks (ER networks, Regular networks, Watts–Strogatz model (WS) networks \cite{watts1998collective}, and BA networks) * 3 mean degrees ($\langle k \rangle = 4, 6, 8$, respectively) * 800 = 9600 instances for training, likewise, 1200 instances for cross-validation, and 1200 data for testing. The epoch of model training is 10, and the batch size is 4. All this data is shuffled for training. The corresponding results are depicted in Figure \ref{fig:model nets}.

Considering potential disparities between synthetic and real-world networks \cite{fan2021characterizing}, we introduced a broader array of empirical networks during the second stage of training to enhance the model's performance in empirical network evaluation scenarios. These empirical networks, diverse in type including social, economic, technical, and communication networks, were sampled to form connected networks of size 1000 and incorporated into the training set, with each type comprising 1000 instances, totaling 5000. Consequently, the second stage introduced an additional 4000 training instances, with 500 instances each for cross-validation and testing. By incorporating instances of these empirical networks into the training process, the model acquired improved generalization capabilities to handle untrained network types, as illustrated in Figure \ref{fig:real1}. Moreover, when the target network aligns with the types used for training, more accurate prediction results are obtained, as shown in Figure \ref{fig:real2}.

The experiments are performed on a PC with a 64-bit Windows 11 Operating System, installed with an Intel (R) 16-Core i7-11700F (2.50GHz) CPU.

\section{Results}\label{results}

\begin{figure}[h]
\centering
\includegraphics[width=\linewidth]{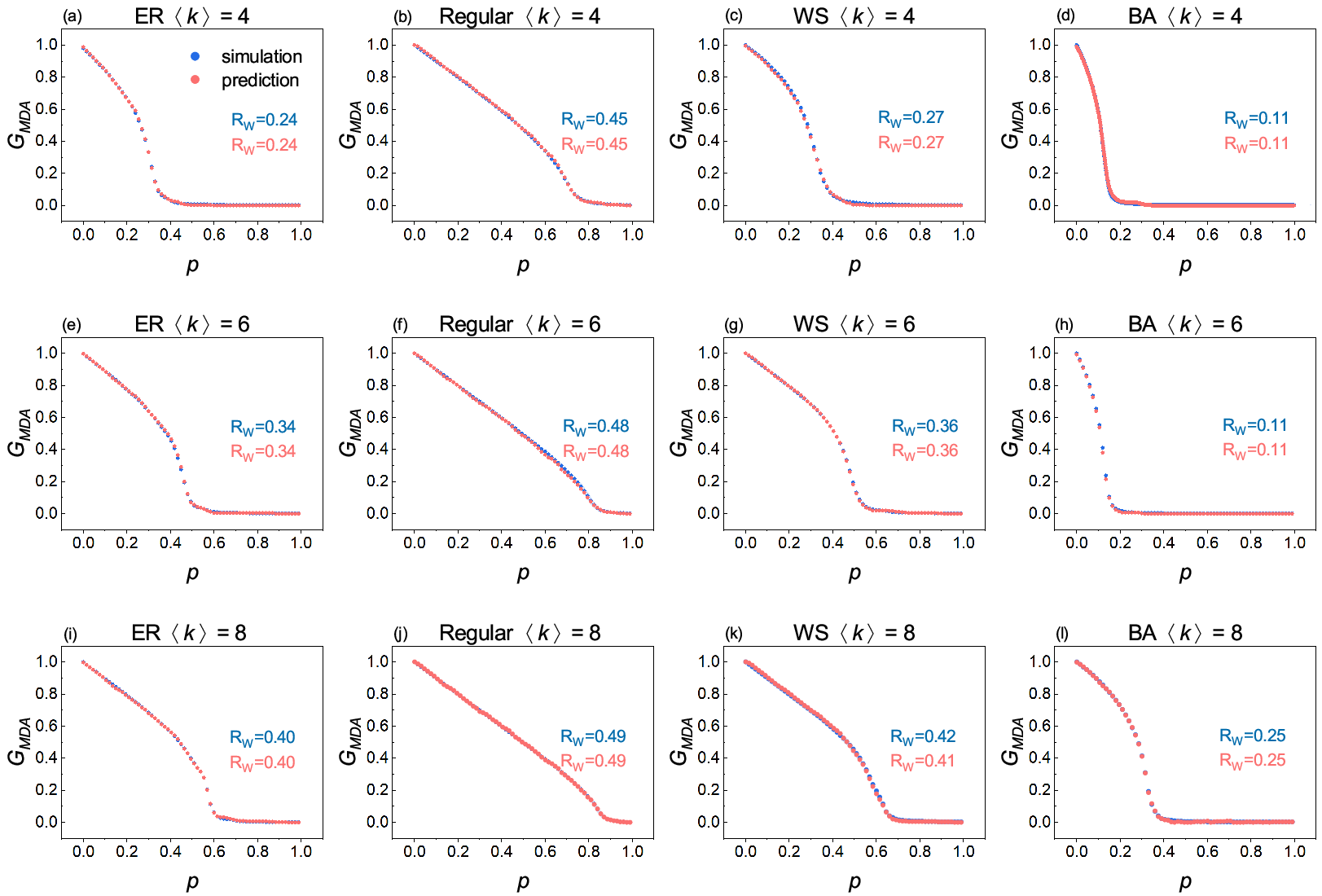}
\caption{Comparison of simulated and the quick evaluator predicted MDA curves on four types of synthetic networks. Here each panel title specifies the network type and average degree, $p$ and $G_{MDA}$ denote the node removal ratio and the corresponding relative size of the GCC under MDA, respectively. Blue and red dots represent simulated and Quick Evaluator predicted results, while blue and red $R_W$ denote the worst robustness values obtained based on the response curves. It is noteworthy that each point represents the average result of 100 independent experiments, and given their standard deviations are very small, they are indistinguishable in these panels.}
	\label{fig:model nets}
\end{figure}

To assess the performance of the proposed WRE framework, we conducted tests on various synthetic and empirical networks, demonstrating the framework's outstanding performance with rapid and accurate results. Figure \ref{fig:model nets} illustrates the evaluation results of synthetic networks, including ER networks, Regular networks, WS networks, and BA networks. It is evident that the predicted MDA curves closely match the simulation results, and the consistency of the two $R_W$ values further quantitatively confirms the excellent performance of the Quick Evaluator. Additionally, it can be observed that BA networks exhibit the smallest worst robustness, with their MDA curves declining rapidly in the early stages. Conversely, Regular networks demonstrate the largest worst robustness, with a more gradual decline in their MDA curves. ER networks and WS networks fall between these extremes. Furthermore, for each column, as the average degree increases, the worst robustness values of all types of networks also increase, indicating that increased density effectively enhances their resilience against various attacks.

\begin{figure}[htbp]
\centering
\includegraphics[width=0.6\linewidth]{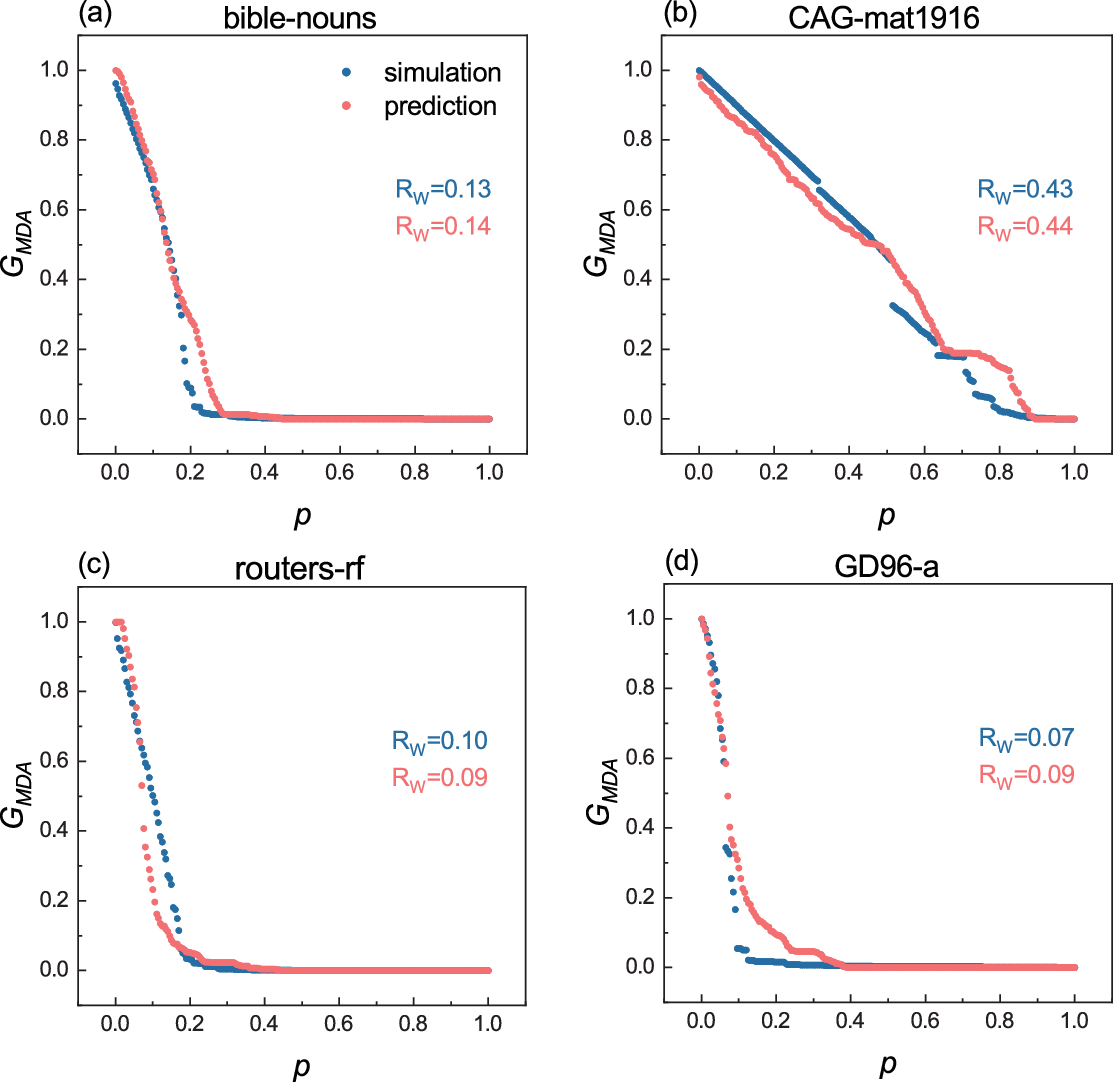}
\caption{Performance of the proposed quick evaluator in predicting empirical networks when the predicted network types differ from those used in training. Here each panel's title specifies the name of the empirical network, $p$ and $G_{MDA}$ denote the node removal ratio and the corresponding relative size of the GCC under MDA, respectively. Blue and red dots represent simulated and Quick Evaluator predicted results, while blue and red $R_W$ denote the worst robustness values obtained based on the response curves.}
	\label{fig:real1}
\end{figure}

Figure \ref{fig:real1} illustrates the performance of the proposed WRE framework in assessing the worst robustness of empirical networks. These networks include the bible-nouns network \cite{kunegis2013konect} which had assembled a digital dataset of cross references found in the King James Bible, the CAG-mat1916 network \cite{nr}, a combinatorial problem network, routers-rf network which is related to routing radio frequency, and the miscellaneous GD96-a network \cite{nr}. These networks exhibit significant differences in their topological characteristics, and their domains are not present in the training set, reflecting the generalization ability of this evaluation framework. The detailed attributes of these networks are listed under Type I in Table \ref{tab:realnets}. 

\begin{table}[htbp]
\caption{Basic topological features of the eight real-world networks. Here $N$, $M$, $\langle k \rangle$ and $\langle L \rangle$ are numbers of nodes, numbers of edges, mean degree, and average shortest path length of the network, respectively. In Type I, the domains of these empirical networks have not appeared in the training dataset, while in Type II, their domains are included in the training dataset.}
\centering
\setlength{\tabcolsep}{5mm}{}
\begin{tabular}{clccccc}
\hline
\multicolumn{2}{c}{Networks} & $N$  & $M$  & $\langle k\rangle$ & $\langle L\rangle$ \\ \hline
\multicolumn{1}{c}{\multirow{4}{*}{Type \uppercase\expandafter{\romannumeral1}}} &   bible-nouns     & 1771       & 9131       & 10.3  &3.3   \\ 
\multicolumn{1}{c}{} &   GD96-a     & 1096   & 1677       & 3.1  & 5.7 \\ 
\multicolumn{1}{c}{} & CAG-mat1916     & 1916    & 164934  & 172.2 & 2.3   \\ 
\multicolumn{1}{c}{} & routers-rf        & 2113      & 6632       & 6.3  &4.6 \\ \hline
\multicolumn{1}{c}{\multirow{4}{*}{Type \uppercase\expandafter{\romannumeral2}}} & tech-as-caida2007     & 1000    & 3342  & 6.7 & 3.7   \\ 
\multicolumn{1}{c}{} & fb-pages-company     & 1000    & 2490  & 5.0 & 6.1   \\ 
\multicolumn{1}{c}{} & econ-poli-large     & 1000    & 1173  & 2.3 & 7.3   \\ 
\multicolumn{1}{c}{} & email-enron-large     & 1000    & 6119  & 12.2 & 3.4   \\ 
\hline
\end{tabular}
\label{tab:realnets}
\end{table}

It can be observed that for these networks, our evaluator accurately predicts the trend of their MDA curves, although there may be slight discrepancies in local details. These discrepancies often occur at positions where the curve trend changes abruptly, indicating significant variations in connectivity resulting from the removal of individual nodes, making precise predictions extremely challenging. However, these deviations in fluctuations may offset each other in the computation of the worst robustness $R_W$, ultimately resulting in only minor differences between predicted and simulated $R_W$, which is deemed acceptable in practice. These results demonstrate that our evaluator can assess the worst robustness of untrained networks relatively accurately, indicating the transferability of the model's predictive capabilities. 

Furthermore, similar to the results for model networks, when the network is sparse, its MDA curve rapidly declines in the early stages of the attack, indicating very low lower bounds on the robustness of these networks, implying their profound vulnerability; whereas for dense networks, such as the CAG-mat1916 network, the decline is gradual, demonstrating strong resistance to various attacks.

\begin{figure}[htbp]
\centering
\includegraphics[width=0.6\linewidth]{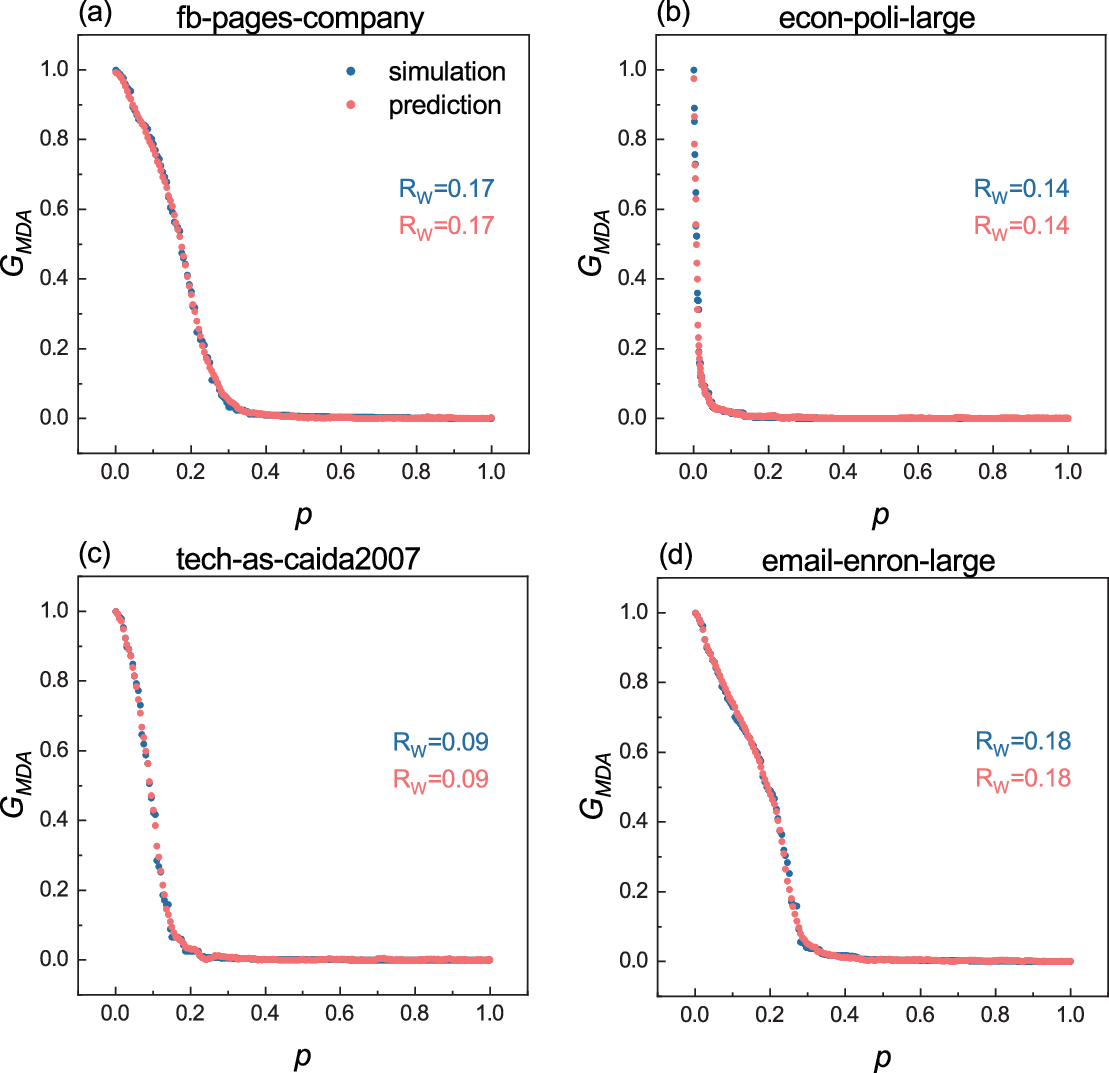}
\caption{Performance of the proposed quick evaluator in predicting empirical networks when the predicted network type matches the training network type. Here each panel's title specifies the name of the empirical network, $p$ and $G_{MDA}$ denote the node removal ratio and the corresponding relative size of the GCC under MDA, respectively. Blue and red dots represent simulated and Quick Evaluator predicted results, while blue and red $R_W$ denote the worst robustness values obtained based on the response curves.}
	\label{fig:real2}
\end{figure}

Figure \ref{fig:real2} presents the evaluation results of four additional empirical networks whose types are encompassed by the training data. These networks include a technical network, tech-as-caida2007 \cite{nr}, a social network, fb-pages-company \cite{nr}, an economic network, econ-poli-large \cite{nr}, and an email communication network, email-enron-large \cite{nr}, whose detailed topological characteristics are listed under Type II in Table \ref{tab:realnets}. Compared to Figure \ref{fig:real1}, it can be observed that both the MDA curves and the worst robustness values $R_W$ of these networks' predicted results are nearly perfectly consistent with the simulated results. These results demonstrate that when the training is sufficiently comprehensive and thorough, the predicted results are able to accurately forecast the simulation results, regardless of whether these networks are robust or fragile.

\section{Discussion}\label{discussion}
This paper addresses the problem of network worst robustness, an important yet often overlooked feature closely related to system security and resilience in practice. By defining the Maximum Damage Attack on a network, we conceptualize Worst Robustness of a network and develop a feasible worst robustness capture scheme based on stacked strategies. Finally, we introduce a deep learning model to address the high time-consuming problem of this evaluation process. We argue for the rationality of the WRE framework, and comprehensive results further demonstrate its outstanding performance. Worst Robustness feature can provide practical solutions for the design, maintenance, and enhancement of resilience in networked systems, particularly for addressing defense issues in networks facing severe malicious attacks, such as military networks and critical infrastructure networks, with potential applications.

The WRE framework exhibits good scalability and can be further improved in its performance. On the one hand, considering the incorporation of more effective attack strategies, or their adaptive versions (which are static in the text), can further enhance the accuracy of this framework to make it closer to theoretical optimality. On the other hand, finding alternative models with stronger predictive capabilities during the prediction phase can further improve the accuracy of predictions.

Furthermore, besides being used for the worst robustness evaluation, the MDA also provides the best-performing attack strategy. This paradigm cannot be provided independently by any known theory but undoubtedly possesses optimal attack performance. Moreover, this inductive paradigm also opens up a new avenue for node ranking and optimal control robustness research, with the potential to confidently expect better-performing critical node identification and ranking schemes.

\section*{Methods}
\label{sec:methods}

\subsection*{1. Node Centralities}
\label{sec:methods_1}
\vspace{0.5\baselineskip}

\noindent \textbf{Degree} In an simple undirected network $G(N, M)$ consisting of $\left| N \right|$ nodes and $\left| M \right|$ edges, the degree of a node $i$ is defined as $d(i)=\sum_{j}a_{ij}$, where $a_{ij}$ denotes the elements of the adjacency matrix of $G$. If two nodes $i$ and $j$ are connected, $a_{ij}=1$; otherwise, $a_{ij}=0$. This measure suggests that a node with more neighbors is more important, and while it is simple and useful in some cases \cite{iyer2013attack}, it overlooks the node's position and the quality of its neighbors.

\vspace{0.5\baselineskip}

\noindent \textbf{H-index} The H-index of a node is defined as the largest $h$ for which it has at least $h$ neighbors, each with a degree no less than $h$ \cite{lu2016vital,korn2009lobby}.
\vspace{0.5\baselineskip}

\noindent \textbf{Coreness} Coreness (or $k$-core) is able to assess the importance of nodes by their location in the network \cite{kitsak2010identification}. Initially, nodes with a degree of $k = 1$ are systematically removed from the network. This process continues until only nodes with a degree of $k > 1$ remain, forming the 1-shell with a coreness of $c = 1$. Subsequently, this method is applied to degrees $k = 2, ... m$, where $m \leq k_{max}$ and $k_{max}$ is the maximum degree. The decomposition process concludes when all nodes have been removed. Ultimately, the coreness $c$ for each node can be determined.

\vspace{0.5\baselineskip}
\noindent \textbf{Closeness} Closeness measures a node's ability to quickly access and disseminate information within a network. Nodes with higher scores are considered more central. The calculation of Closeness centrality is as follows:

\begin{equation} \label{eq:CC}
CC(i)=\frac{1}{N(N-1)}\sum_{i=1,j=1}\sum_{j \neq i}\frac{1}{d_{ij}},
\end{equation}

\vspace{0.5\baselineskip}
\noindent where $CC(i)$ represents the Closeness centrality of node $i$, $N$ denotes the number of nodes in the network, $d_{ij}$ represents the shortest path distance between nodes $i$ and $j$. 

\vspace{0.5\baselineskip}
\noindent \textbf{Betweenness} Betweenness \cite{freeman1977set,sabidussi1966centrality} quantifies the number of shortest paths that pass through a node, highlighting its role as a bridge or mediator in the network. Nodes with high betweenness centrality are crucial for facilitating communication and information flow between different parts of the network. It defined as follows:

\begin{equation}
    BC(i) = \sum_{s \neq i \neq t} \frac{\sigma_{st}(i)}{\sigma_{st}},
\end{equation}

\vspace{0.5\baselineskip}
\noindent where $\sigma_{st}$ represents the total number of shortest paths from node $s$ to node $t$, and $\sigma_{st}(i)$ represents the number of shortest paths from node $s$ to node $t$ that pass through node $i$.

\vspace{0.5\baselineskip}
\noindent \textbf {Eigenvector} Eigenvector \cite{bonacich1987power} considers not only a node's degree but also the degrees of its neighbors. It assigns greater centrality scores to nodes connected to other highly central nodes. It can be obtained using the following equation:

\begin{equation}
    EC(i) = \frac{1}{\lambda} \sum_{j} A_{ij} \cdot EC(j),
\end{equation}

\vspace{0.5\baselineskip}
\noindent where $A_{ij}$ represents the element of the adjacency matrix corresponding to the connection between node $i$, and node $j$, and $\lambda$ represents the dominant eigenvalue of the adjacency matrix.

\vspace{0.5\baselineskip}
\noindent \textbf{Pagerank} PageRank \cite{brin1998anatomy} assigns importance scores to nodes based on the structure of the network and the quality of their connections. It is defined as follows:
\begin{equation}
PR(i) = \frac{1 - d}{N} + d \sum_{j \in M(i)} \frac{PR(j)}{L(j)},
\end{equation}
where $d$ is the damping factor, $N$ is the total number of nodes in the network, $M(i)$ represents the set of nodes that link to node $i$, and $L(j)$ denotes the number of outgoing edges from $j$.

\vspace{0.5\baselineskip}
\noindent \textbf{Cycle ratio} Cycle Ratio measures the importance of individual nodes based on their involvement in cyclic structures within the network \cite{fan2021characterizing}. It defined as follows:
\begin{equation}
    r_i = \left\{ \begin{aligned}
  & 0,c_{ii}=0 \\ 
 & \sum_{j,c_{ij}>0}\frac{c_{ij}}{c_{jj}}, c_{ii}>0,\\ 
\end{aligned} \right.
\end{equation}
In network $G$, $S$ is the set of all shortest basic cycles of $G$, $c_{ij}$ is the number of cycles in $S$ that pass through both nodes $i$ and $j$ if $i \neq j$, $c_{ii}$ is the number of cycles in $S$ that contain node $i$.

\subsection*{2. Maximum Rationality of MDA}

The Maximum Rationality (MR) of the MDA curve refers to the maximum proportion of unique nodes counted only once along the MDA curve. For instance, in a network of size 1000, if the MDA curve contains 995 unique nodes, then its MR is 0.995. The MDA curve is constructed by stacking multiple strategies, where at each position, there might be multiple nodes contributing the same maximum damage, or a single node appearing multiple times at different positions. In such cases, arranging these nodes strategically can maximize the MR. If the adjusted MDA curve's MR equals 1, then the removal sequence represented by this MDA curve is absolutely rational, indicating that no nodes remain in the network after all nodes have been removed once.

Given $q$ candidate attack curves $S_1, S_2, \ldots, S_i, \ldots, S_q $, where $S_i=[(r_1, n_1)$, $(r_2, n_2), \ldots, (r_N, n_N)] $, with $ r_1, r_2, \ldots, r_N $ representing the sequence of Robustness values obtained after applying strategy $ i $, and $ n_1, n_2, \ldots, n_N $ denoting the corresponding node removal sequence, where $ N $ denotes the number of nodes. Thus, each node has $ q $ values of $ r $. The process of capturing the MDA curve and calculating its Maximum Rationality (MR) is as follows:

\begin{enumerate}
    \item Initially, we assign a counter for each node and initialize all counters to 0, $ \{c_i\} = 0 $, where $ c_i $ represents the number of occurrences of node $ i $ in the MDA curve.
    
    \item Next, we iteratively insert the maximum damage value for the current step into one of the $ N $ positions of the MDA curve, updating the counters for the corresponding nodes. At position $ p_j $, if only one node among the $ q $ strategies has the minimum $ r_j $ value, then $ r_j $ is inserted at position $ p_j $ of the MDA curve, and the counter for the corresponding node is incremented by 1. If multiple nodes have the same minimum $ r_j $, the node with the smallest counter is chosen, and its counter is incremented by 1. This process continues until all $ N $ positions have been processed.
    
    \item Subsequently, we further optimize the counters for all nodes to minimize the cases where counters are equal to 0 or greater than 1. If a node's counter is 0 and one of its $ r $ values equals the minimum robustness value at position $ p_j $ of the MDA curve, then the original node is replaced with this node, the counter of this node is incremented by 1, and the counter of the original node is decremented by 1. This operation iterates until no more nodes can be replaced at any position.
    
    \item Finally, if the number of nodes with a counter equal to 0 is $ u_0 $, then the Maximum Rationality of this MDA curve is calculated as $ MR = (N - u_0)/N $.
\end{enumerate}

\section*{CRediT authorship contribution statement}
Wenjun Jiang, Tianlong Fan, and Zong-fu Luo conceived the idea and designed the study. Tianlong Fan and Zong-fu Luo managed the study. Wenjun Jiang and Peiyan Li collected and cleaned up the data. Wenjun Jiang, Tianlong Fan, and Peiyan Li performed the experiments. Wenjun Jiang and Tianlong Fan wrote the manuscript, Ting Li supplemented visualization. Tianlong Fan, Zong-fu Luo, Tao Zhang, and Chuanfu Zhang edited this manuscript. All authors discussed the results and reviewed the manuscript.
\section*{Declaration of completing interest}
The authors declare no competing interests.

\section*{Acknowledgments}
This work was supported in part by the National Natural Science Foundation of China (Grant No. T2293771), the National Laboratory of Space Intelligent Control (No. HTKJ2023KL502003), the Fundamental Research Funds for the Central Universities, Sun Yat-sen University (No. 23QNPY78) and the STI 2030--Major Projects (No. 2022ZD0211400).




\journal{Journal of \LaTeX\ Templates}
 \bibliographystyle{elsarticle-num} 
\bibliography{reference}

\newpage
\addcontentsline{toc}{section}{Appendix}
\section*{Appendix}\label{sec:appendix}
\setcounter{figure}{0}
\setcounter{table}{0}

\renewcommand{\thefigure}{\thesection.\arabic{figure}}
\renewcommand{\thefigure}{A\arabic{figure}} 
\begin{figure} [htbp]
\centering
\makebox[\textwidth][c]{\includegraphics[width=1.29\textwidth]{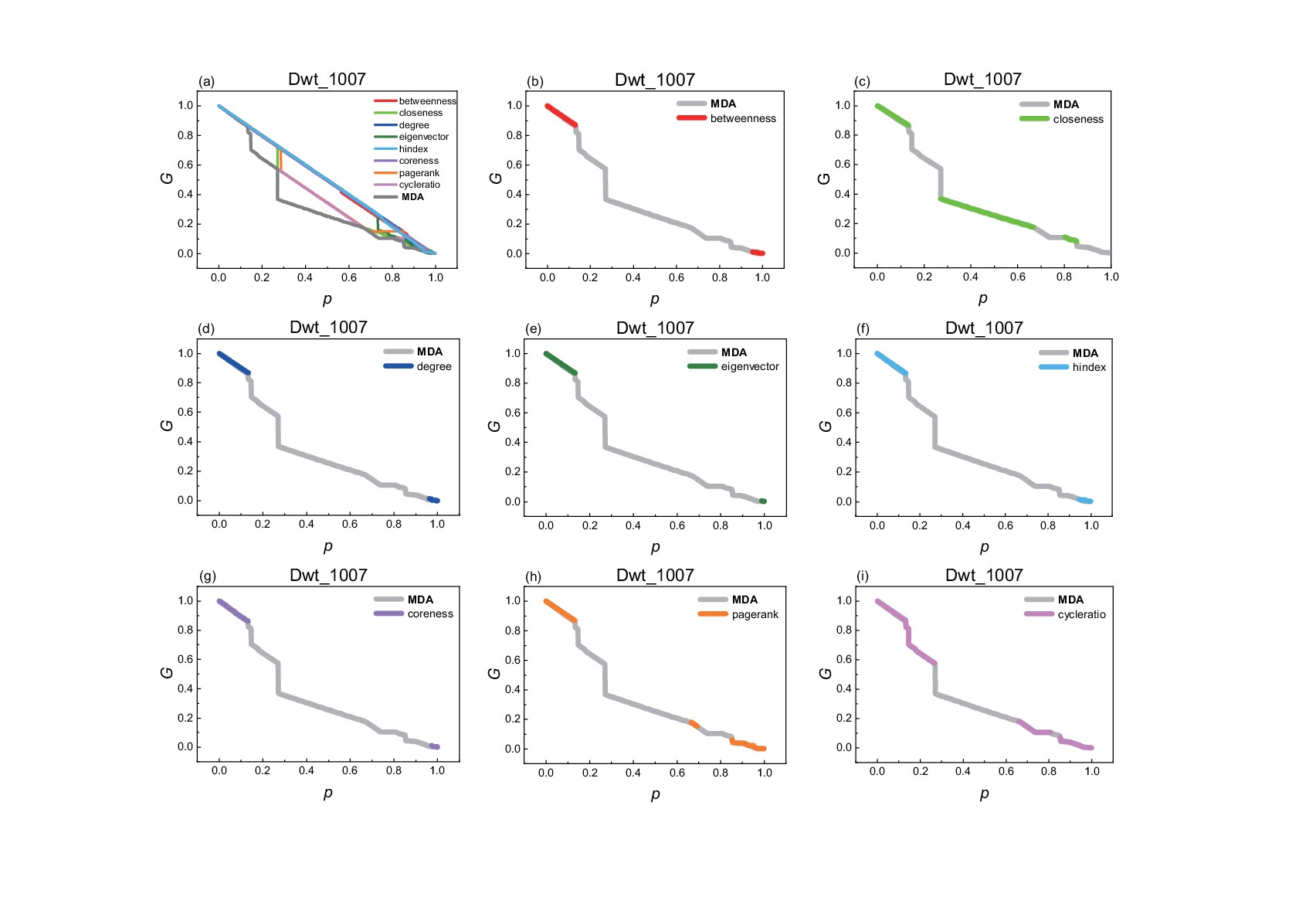}}
\caption{The different attack curves for the Dwt\_1007 network and the decomposition schematic of the worst robustness curve are in the residual panels. $p$ and $G$ represent the proportion of removed nodes and the relative size of the GCC respectively.
The Dwt\_1007 network is an undirected and unweighted miscellaneous network, consisting of 1007 nodes and 3784 edges \cite{nr}.}
	\label{fig:sample}
\end{figure}


\renewcommand{\thetable}{\thesection.\arabic{table}}
\renewcommand{\thetable}{A\arabic{table}}

\begin{table}[]
\caption{Statistics of the maximum rationality in the synthetic networks with eight centrality metrics. For each set of parameters, we randomly generated 100 networks. The first three columns display their Maximum (Max), Minimum (Min), and Mean MR, while the remaining elements indicate the number of networks satisfying the right-tail distribution condition for MR.}
\centering
\small
\begin{tabular}{cl|c|c|l|c|c|c|c|c|c}
\hline
\multicolumn{2}{c|}{Networks}                         & Max & Min & Mean & \textgreater{}0.95 & \textgreater{}0.9 & \textgreater{}0.85 & \textgreater{}0.80 & \textgreater{}0.75 & \textgreater{}0.7 \\ \hline
\multicolumn{1}{c|}{\multirow{3}{*}{BA}}      & $\langle k\rangle =4$ & 0.99    & 0.94    & 0.97 & 98                 & 2                 &                    &                    &                    &                   \\
\multicolumn{1}{c|}{}                         & $\langle k\rangle =6$ & 0.99    & 0.95    & 0.98 & 99                 & 1                 &                    &                    &                    &                   \\
\multicolumn{1}{c|}{}                         & $\langle k\rangle =8$ & 0.99    & 0.95    & 0.98 & 100                &                   &                    &                    &                    &                   \\ \hline
\multicolumn{1}{c|}{\multirow{3}{*}{ER}}      &  $\langle k\rangle =4$  & 1.00       & 0.93    & 0.98 & 97                 & 3                 &                    &                    &                    &                   \\
\multicolumn{1}{c|}{}                         & $\langle k\rangle =6$ & 1.00       & 0.94    & 0.99 & 99                 & 1                 &                    &                    &                    &                   \\
\multicolumn{1}{c|}{}                         & $\langle k\rangle =8 $ & 1.00       & 0.92    & 0.98 & 90                 & 10                &                    &                    &                    &                   \\ \hline
\multicolumn{1}{c|}{\multirow{3}{*}{Regular}} &  $\langle k\rangle =4$  & 1.00       & 0.91    & 0.97 & 79                 & 21                &                    &                    &                    &                   \\
\multicolumn{1}{c|}{}                         & $\langle k\rangle =6$ & 1.00       & 0.91    & 0.97 & 83                 & 17                &                    &                    &                    &                   \\
\multicolumn{1}{c|}{}                         & $\langle k\rangle =8 $ & 1.00       & 0.90    & 0.97 & 86                 & 14                &                    &                    &                    &                   \\ \hline
\end{tabular}
\label{tab:A1}
\end{table}

\begin{table}[]
\caption{Statistics of the maximum rationality in the synthetic networks with twelve centrality metrics. The first three columns display their Maximum (Max), Minimum (Min), and Mean MR, while the remaining elements indicate the number of networks satisfying the right-tail distribution condition for MR. The other four metrics are HITS (Hyperlink-induced topic search) \cite{kleinberg1999authoritative}, Subgraph centrality \cite{estrada2005subgraph}, load centrality \cite{newman2001scientific} and CI (Collective influence) \cite{morone2015influence}.}
\centering
\small
\begin{tabular}{cl|c|c|l|c|c|c|c|c|c}
\hline
\multicolumn{2}{c|}{Networks}                         & Max & Min & Mean & \textgreater{}0.95 & \textgreater{}0.9 & \textgreater{}0.85 & \textgreater{}0.80 & \textgreater{}0.75 & \textgreater{}0.7 \\ \hline
\multicolumn{1}{c|}{\multirow{3}{*}{BA}}      & $\langle k\rangle =4$  & 0.99    & 0.93    & 0.97 & 98                 & 2                 &                    &                    &                    &                   \\
\multicolumn{1}{c|}{}                         & $\langle k\rangle =6$ & 0.99    & 0.95    & 0.97 & 99                 & 1                 &                    &                    &                    &                   \\
\multicolumn{1}{c|}{}                         & $\langle k\rangle =8 $ & 0.99    & 0.95    & 0.98 & 100                &                   &                    &                    &                    &                   \\ \hline
\multicolumn{1}{c|}{\multirow{3}{*}{ER}}      & $\langle k\rangle =4$  & 0.92    & 0.86    & 0.89 &                    & 38                & 62                 &                    &                    &                   \\
\multicolumn{1}{c|}{}                         & $\langle k\rangle =6$ & 0.87    & 0.83    & 0.86 &                    &                   & 95                 & 5                  &                    &                   \\
\multicolumn{1}{c|}{}                         & $\langle k\rangle =8 $ & 0.86    & 0.83    & 0.84 &                    &                   & 7                  & 93                 &                    &                   \\ \hline
\multicolumn{1}{c|}{\multirow{3}{*}{Regular}} & $\langle k\rangle =4$  & 0.85    & 0.75    & 0.77 &                    &                   &                    & 4                  & 93                 & 3                 \\
\multicolumn{1}{c|}{}                         & $\langle k\rangle =6$ & 0.90    & 0.77    & 0.82 &                    & 1                 & 9                  & 76                 & 14                 &                   \\
\multicolumn{1}{c|}{}                         & $\langle k\rangle =8 $ & 0.91    & 0.80    & 0.86 &                    & 4                 & 55                 & 40                 & 1                  &                   \\ \hline
\end{tabular}
\label{tab:A2}
\end{table}

\begin{figure} [htbp]
\centering
\includegraphics[width=0.8\linewidth]{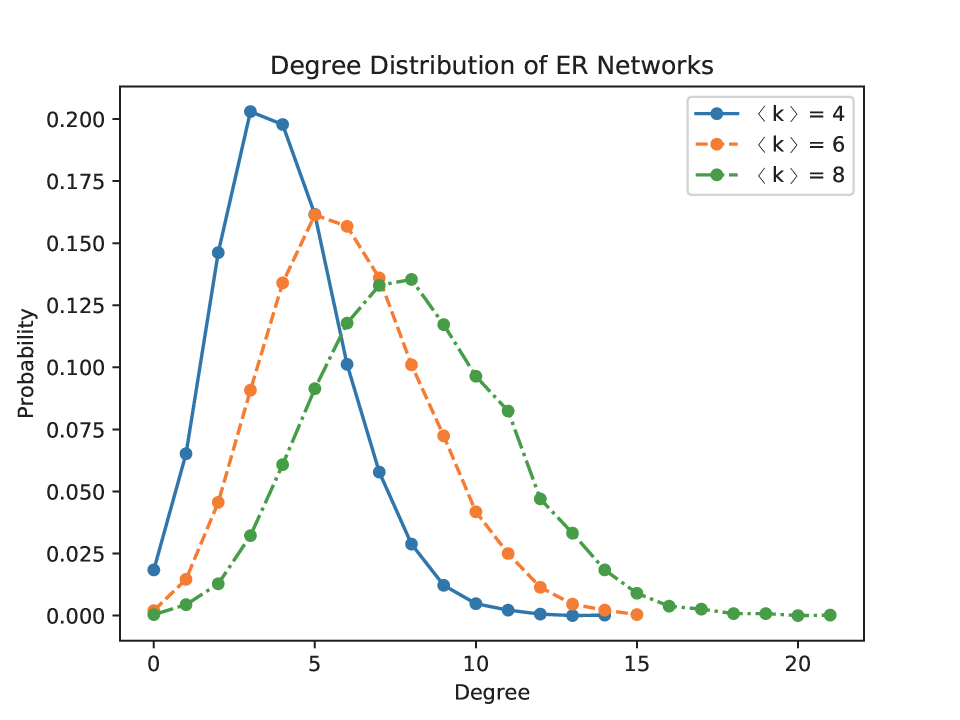}
\caption{Comparison of degree distributions in ER networks with different average degrees. Here all three networks consist of 5000 nodes.}
	\label{fig:ERdistr}
\end{figure}

\end{document}